\documentclass[11pt]{article}
\usepackage{amsmath}
\usepackage{graphicx,psfrag,epsf,float, bm, subcaption, url, makecell,authblk, adjustbox}
\usepackage{enumerate}
\usepackage{natbib, booktabs}
\usepackage{undertilde}
\usepackage{url} 
\graphicspath{{figures_tables/}}

\usepackage[x11names]{xcolor}
\usepackage{framed} 
\definecolor{shadecolor}{rgb}{1,.5,.5}

\newcommand{\ind}{\stackrel{ind.}{\sim}}
\newcommand{\op}{\operatorname}

\newcommand{\myequation}{\begin{equation}}
\newcommand{\myendequation}{\end{equation}}
\let\[\myequation
\let\]\myendequation

\newcommand{\splusbookfiguresizexx}[2]{
\includegraphics[keepaspectratio=true,width=#2,angle=0]{#1}
}

\newcommand{\blind}{0}

\begin{document}
\def\spacingset#1{\renewcommand{\baselinestretch}%
{#1}\small\normalsize} \spacingset{1}


\if0\blind
{
  \title{\bf Prediction of Future Failures for Heterogeneous Reliability Field Data}
\author[1]{Colin Lewis-Beck}
\author[2]{Qinglong Tian}
\author[2]{William Q. Meeker}
\affil[1]{Department of Statistics \& Actuarial Science, University of Iowa}
\affil[2]{Department of Statistics, Iowa State University}
  \maketitle
} \fi

\if1\blind
{
  \bigskip
  \bigskip
  \bigskip
  \begin{center}
    {\LARGE\bf Prediction of Future Failures for Heterogenous Reliability Field Data}
\end{center}

  \medskip
} \fi

\begin{abstract}

This article introduces methods for constructing prediction bounds or intervals for the number of future failures from heterogeneous reliability field data.  We focus on within-sample prediction where early data from a failure-time process is used to predict future failures from the same process.  Early data from high-reliability products, however, often have limited information due to some combination of small sample sizes, censoring, and truncation. In such cases, we use a Bayesian hierarchical model to model jointly multiple lifetime distributions arising from different subpopulations of similar products.  By borrowing information across subpopulations, our method enables stable estimation and the computation of corresponding prediction intervals, even in cases where there are few observed failures. Three applications are provided to illustrate this methodology, and a simulation study is used to validate the coverage performance of the prediction intervals.

\end{abstract}


\noindent%

\spacingset{1.45} 

\section{Introduction}
\label{sec:intro}
Consumers and producers often have to predict future values of random quantities given data from a failure-time process.  These random quantities fall into one of two different classes of prediction problems, referred to as new-sample and within-sample prediction.  New-sample prediction includes making predictions on a future unit or collection of units from the same population.  Within-sample prediction problems predict future events based on early observed data from a failure-time process. Such applications include
\begin{itemize}

\item Managers of a fleet of systems want to predict the number of needed replacement components or subsystems for spare part provisioning.
\item A company needs to predict the number of future failures for a manufactured product to determine if a recall is required.
\item A reliability engineer conducting a life test needs to estimate the number of future failures from a group of products that will occur in a given interval in order to determine when the test will end.
\end{itemize}

In within-sample prediction, computing prediction intervals or one-sided prediction bounds, however, can be challenging. High-reliability products often lead to heavy right censoring.  With few observed failures, relying on large-sample theory for prediction intervals is problematic.   Field data can also involve a population of products with heterogeneous subpopulations.  These subpopulations may enter service over a period of time (known as staggered entry), exhibit different lifetime distributions (e.g., because of different environments or use rates), and have varying amounts of time in service. In this scenario, it may be possible to make predictions only for subpopulations with many observed failures. Jointly modeling the subpopulations, however, provides prediction intervals for the individual subpopulations or the entire population, providing an appealing alternative.

This paper presents a general framework for making within-sample predictions for the number of future failures based on failure-time data from a potentially heterogeneous population of components or subsystems operating within a larger population or fleet of systems. Applications include failure-time data, warranty data, and data with dynamic covariate information. When the number of failures is small, we take a hierarchical modeling approach in order to partially pool subpopulations and reduce the amount of data required to produce stable prediction intervals.  Estimation is performed using Bayesian methods, which allows for prior information about lifetime distribution parameters (if available) to be incorporated into the model.  Moreover, the construction of prediction intervals is relatively straightforward with the Bayesian framework, once the set of draws from the joint posterior distribution are available. 

There are many books and papers on the subject of prediction intervals. \cite{guttman} describes both Bayesian and non-Bayesian tolerance and prediction intervals. \cite{geisser1993predictive} provides an introduction to Bayesian methods for prediction. \citet[Chapter 12]{meeker} present a general overview of reliability-related applications of prediction intervals. \cite{escobar1999statistical} develop a simulation-based method to provide prediction intervals for new and within-sample predictions.  \cite{nelson2000weibull} uses likelihood ratio based prediction intervals for future within-sample failures using a Weibull distribution.  \cite{nordman2002weibull} extend the work of Nelson and study the properties of likelihood ratio-based prediction intervals. 

In other related literature, \cite{hamada2004bayesian} explain the differences between Bayesian tolerance and new-sample prediction intervals and illustrate these with an example using a hierarchical linear model. \cite{hong2009prediction} develop methods to predict the number of future failures from a fleet of power transformers that exhibit both left truncation and right-censoring. Other applications include \cite{wang2009life}, who predict the future life of a series system comprised of exponential lifetime distributions, and study the frequentist coverage probabilities of their Bayesian intervals. \cite{hong2010field} predict warranty returns for a product with multiple failure modes.  \cite{hong2013field} later extend this work by incorporating dynamic covariate information to make field-failure predictions both for individual units and for a population. 

The contributions of this article are summarized as follows.  We develop a general method to make within-sample predictions for heterogeneous reliability field data using Bayesian methods.  Our procedures can be applied to data with various censoring and truncation, limited observed failures, dynamic risk sets, as well as fixed and time-varying covariates.  We illustrate and validate our methods using field data from three different types of products.  We also demonstrate, through a simulation study, the potential advantages of using hierarchical models to borrow information across subpopulations to improve prediction intervals.

The remainder of this paper is organized as follows. Section \ref{sec:Data} introduces three motivating applications for our prediction methodology. Section \ref{sec:weib model} describes general hierarchical lifetime models that can include varying types of censoring and truncation commonly encountered in reliability field data.  Section \ref{sec:predfail} presents methodology for constructing prediction intervals for the number of future failures from subpopulations, as well as an entire population.  Section \ref{sec:application} applies our methods to the motivating applications. Section \ref{sec:sim} describes a small simulation study to study the coverage probability properties of our Bayesian prediction intervals based on weakly informative prior distributions. Section \ref{sec:conclusion} contains some concluding remarks and suggestions for future research.

\section{Motivating Applications}
\label{sec:Data}
\subsection{Heat Exchanger Tube Data}
Nuclear power plants contain steam generators which are large heat exchangers comprised of many stainless steel tubes. Over time, the tubes can crack due to a combination of cyclic stressing and corrosion. Periodic inspection is used to detect such cracks. Once a crack is identified, it is plugged and the tube is thus removed from service. Given the observed failures, engineers would like to predict the number of additional tubes, $Y$, that will crack over the next $t_c + \Delta t$ years. Both \cite{nelson2000weibull} and  \cite{nordman2002weibull} analyzed similar data using maximum likelihood (ML) estimation, and predicted the number of additional tubes that would fail over the next 10 years.  However, due to a limited number of observed failures, data from many plants were pooled together to improve estimation.  In addition, the value of the Weibull shape parameter, $\beta$, was assumed to be known and the same in all of the plants.


\subsection{Backblaze Hard Drive Data}
\label{sec:bbdata}
Backblaze is a company that provides cloud backup storage to protect against customer data loss.  Since 2013, Backblaze has been collecting daily operational data on all of the hard drives operating at its facilities.  Every quarter the company reports detailed operational data and summary statistics on the different drive-models in operation through their website (\url{https://www.backblaze.com/b2/hard-drive-test-data.html}, accessed June 1, 2020)\nocite{backblaze}.  The purpose is to provide consumers and businesses with reliability information on different drive-models.  The hard drives continuously spin in controlled-environment storage pods.  Drives are run until failure or until they are replaced with newer technology drives.  When a hard drive fails, it is removed and replaced.  In addition, the number of storage pods is increasing as Backblaze expands their business and adds drives to its storage capacity. A subset of these data was analyzed by \cite{mittman}.  However, their focus was on comparing the reliability of the different drive-model brands whereas our interest is in predicting the number of future failures over a fixed future period of time for a current population of drives.


\subsection{Product A Data}
\label{sec:proda}
Monthly predictions of warranty returns are needed by manufacturing companies for spare part provisioning, pricing of warranty plans, and allocating sufficient warranty cash reserves.  To predict the number of monthly warranty returns accurately, however, it is important to consider heterogeneity in product usage rates. For example, some products are used more in the summer months or fail more frequently in certain geographic regions. Our third application, referred to as Product A, uses warranty return data collected on 63,132 systems from 2011 to 2017.  The Product A data set contains information on the start and expiration date of warranty contracts, country (United States or Canada), model type, retail location, and the event date.   \cite{shan2020seasonal} analyzed the Product A systems as recurrent events data.  For our application, however, we turn the data into failure-time data in order to predict the time to the first warranty return.

\section{Models for Heterogeneous Reliability Field Data}
\label{sec:weib model}
The models presented in this paper could employ any of the
commonly used lifetime distributions such as the Weibull, Fr\'echet, or
lognormal (see Chapter 4 of Meeker et al. 1998 for
others). In this article, we use the Weibull and Fr\'echet
distributions, as well as a mixture of Weibull distributions.  These particular distributions were chosen  based on a combination of engineering judgment and how well the models fit the data \citep{nelson2000weibull, mittman}. However, using any other distribution or combination of distributions in the log-location-scale family of distributions is straightforward.

\subsection{The Weibull and Fr\'echet Distributions, and Reparameterization}
\label{sec:Weibull parameterization}
The Weibull cumulative
 distribution function (cdf) is
\begin{align*}
\Pr(T \leq t|\eta,\beta ) &= F(t|\eta,\beta)=1-
\exp \left [-\left (\frac{t}{\eta} \right )^{\beta}
\right ], \quad t > 0,
\end{align*}
where $\beta>0$ is the Weibull shape parameter and $\eta>0$ is a
scale parameter.  
Because $\log(T)$ has a smallest extreme value
distribution (a member of the location-scale family of
distributions), the Weibull cdf can also be written as
\begin{align}
\label{equation:reparm.cdf}
\Pr(T \leq t| \mu,\sigma ) &= F(t| \mu,\sigma)= \Phi_{\textrm{sev}}\left[\frac{\log(t)-\mu}{\sigma}\right], \quad t > 0,
\end{align}
where $\Phi_\textrm{{sev}}(z)=1-\exp[-\exp(z)]$ is the standard smallest extreme value
distribution cdf and
$\mu=\log(\eta)$ and $\sigma=1/\beta$ are, respectively, location
and scale parameters for the distribution of $\log(T)$. One advantage of using this Weibull parameterization is that it corresponds to the parameterization of the other log-location-scale distributions.

Following the suggestions given in \cite{LiMeeker2014} and Section 15.2 of \cite{intervals}, we use an alternative parameterization
where the usual scale parameter $\eta$ is replaced by the $p$ quantile $t_{p}=\eta \left
[-\log(1-p)\right ]^{\sigma}$ (which is also a scale parameter).  This parameterization is important for estimation stability when there is heavy censoring as well as eliciting prior information. Eliciting information is easier for a particular quantile as opposed to the usual scale parameter (which may not be practically relevant).  The cdf for the Fr\'echet distribution is similarly expressed as in (\ref{equation:reparm.cdf}), but replacing the standard smallest extreme value distribution with the standard largest extreme value distribution $\Phi_\textrm{{lev}}(z)=\exp[-\exp(-z)]$.  For a more general discussion of reparameterization and Bayesian inference, see \cite{gelman2004parameterization}.

\subsection{The Hierarchical Failure Population Model}
\label{subsec:hier model}

In some applications, there is a need to model jointly an entire product population consisting of different but similar subpopulations. When there is a limited amount of data in some of the subpopulations, it is useful to model certain subpopulation-specific parameters hierarchically, borrowing strength across subpopulations. This extension can be applied to any log-location-scale distribution (e.g, the Weibull or the Fr\'echet), but for this application, we extend the Weibull lifetime model as follows,
\begin{equation}
T_{ig}|\bm{\theta_g} \ind \op{Weibull}\left(t_{p_g}, \sigma_{g} \right),
\end{equation}
where $g=1,\ldots,G$ indexes the subpopulations.  The likelihood for the Weibull model for all subpopulations is a function of the sets of parameters $\bm{\theta_g} = (t_{p_g}, \sigma_{g})$, one set for each subpopulation, $g$.  Assuming the lifetimes of all units are independent within and across subpopulations, and conditional on fixed values of the parameters, the likelihood for the data, $i=1,\dots,n_g$,  is given by
\begin{equation*}
L(\bm{\Theta})= \prod_{g=1}^{G} \prod_{i=1}^{n_{g}} \left[f(t_{ig};\bm{\theta_g})\right]^{1-C_{ig} - I_{ig} - L_{ig}} \left[1-F(t_{ig};\bm{\theta_g}) \right]^{C_{ig}} \left[F(t_{1ig};\bm{\theta_g}) \right]^{L_{ig}}\left[F(t_{1ig};\bm{\theta_g}) -F(t_{0ig};\bm{\theta_g}) \right]^{I_{ig}}
\end{equation*}
where $t_{ig}$ is the observed failure or survival time of unit $i$ in subpopulation $g$; $t_{0ig}$ and $t_{1ig}$ ($t_{0ig} < t_{1ig}$) are the lower and upper interval-censored times;  $C_{ig}$ is an indicator if unit $i$ in subpopulation $g$ is right-censored; $L_{ig}$ is an indicator if unit $i$ in subpopulation $g$ is left-censored; and $I_{ig}$ is an indicator if unit $i$ in subpopulation $g$ is interval-censored.  

In a hierarchical model, the parameters are modeled as random, varying across different subpopulations.  Thus, to complete the model we specify the following distributions,
$$
\sigma_{g} \ind \op{Lognormal} \left( \eta_{\sigma}, \tau^2_{\sigma} \right) ,\quad
t_{p_g} = \exp\left(\mu_{g} + \sigma_{g}\,\Phi_\textrm{{sev}}^{-1}(p)\right)  \ind \op{Lognormal} \left(\eta_{t_{p}}, \tau^2_{t_{p}}\right).$$

\subsection{The Hierarchical Generalized Limited Failure Population Model}
\label{sec:hier-glfp}
The Generalized Limited Failure Population model (GLFP) from \cite{chan} is a generalization of the Weibull model to accommodate lifetime data with two failure modes. The GLFP accounts for early failures due to manufacturing defects (infant mortality) and, after this `burn-in' period, failures due to prolonged use (wearout).  The GLFP can be seen as a special case of independent competing risks with two causes of failure,  $F_1,F_2$,  with parameters $(t_{p_1},\sigma_1)$ and $(t_{p_2}, \sigma_2)$, respectively. The GFLP model uses the parameter, $\pi$, the population proportion of defective units that are susceptible to both failure modes. When $\pi$ is zero, the GLFP model reduces to a single Weibull distribution. If $T \sim \op{GLFP}(\pi, t_{p_1},\sigma_1,t_{p_2},\sigma_2)$, then the cdf of $T$ is 
$$\Pr(T \le t) = F(t; \pi, t_{p_1},\sigma_1,t_{p_2},\sigma_2) = 1 - (1-\pi\, F_{1}(t))(1 - F_{2}(t)),\, t>0,\, 0 < \pi < 1.$$ and taking the derivative of the cdf gives us the density for the GLFP, 
$ \partial F(t; \bm{\theta}) / \partial t = f(t;\pi, t_{p_1},\sigma_1,t_{p_2},\sigma_2)$.
Here $F_{1}(t)$ is the cdf for the early failures and $F_{2}(t)$ is
the cdf for the wearout failures.

We extend the GLFP model as follows,
\begin{equation}
T_{ig}|\bm{\theta_g} \ind \op{GLFP}\left( \pi_g, t_{p_{1}g}, \sigma_{1g}, t_{p_{2}g}, \sigma_{2g} \right),
\end{equation}
where again $g=1,\ldots,G$ indexes the subpopulations.  We now define $\bm{\theta_g} = (\pi_{g}, t_{p_{1}g}, \sigma_{1g}, t_{p_{2}g}, \sigma_{2g})$, one set for each subpopulation, $g$.  Under the same independence assumptions as for the single Weibull distribution, the likelihood for the data, $i=1,\dots,n_g$, is 
\begin{equation*}
L(\bm{\Theta})= \prod_{g=1}^{G} \prod_{i=1}^{n_{g}} \left[\frac{h(t_{ig};\bm{\theta_g})}{1-H(t_{ig}^L;\bm{\theta_g})}\right]^{1-C_{ig}} \left[ \frac{1-H(t_{ig};\bm{\theta_g})}{1-H(t_{ig}^L;\bm{\theta_g})} \right]^{C_{ig}}
\end{equation*}
where $t_{ig}$ is the observed failure or survival time of unit $i$ in subpopulation $g$; $t_{ig}^L$ is the left truncation time for unit $i$ in subpopulation $g$; and $C_{ig}$ is an indicator if unit $i$ in subpopulation $g$ is right-censored.  As in Section \ref{subsec:hier model}, including left- and interval-censored observations into the likelihood is straightforward.

Assuming a distribution for early failures, common across the $G$ groups, provides a meaningful interpretation and comparison of the proportion defective, $\pi_g$, and typically, the
lifetime distribution of the defective subpopulation is not of high interest in practical applications.  Thus, we only specify hierarchical distributions for the proportion defective and the Weibull parameters for the wearout failure mode,
\begin{equation*}
\sigma_{2g} \ind \op{Lognormal} \left( \eta_{\sigma_2}, \tau^2_{\sigma_2}\right)\op{Tr}\left(0, 1\right), \quad \pi_g \ind \op{Logit-normal}(\eta_\pi, \tau^2_\pi),
\end{equation*}
\begin{equation*}
t_{p_{2}g} = \exp\left(\mu_{2g} + \sigma_{2g}\,\Phi_\textrm{{sev}}^{-1}(p_2)\right)  \ind \op{Lognormal} \left(\eta_{t_{p_2}}, \tau^2_{t_{p_2}}\right).
\end{equation*}
We truncate the distribution of $\sigma_{2g}$ above 1, which restricts the wearout failure mode to have an
increasing hazard function.
\section{Prediction}
\label{sec:predfail}
We focus on the within-sample prediction problem in two contexts: predicting future failures from a single group or subpopulation, and aggregating over all $G$ groups (i.e, the entire population). The aggregation prediction application is somewhat more complicated because the groups of units could differ due to a variety of covariates such as entry time, hazard rates, censoring, or the number of units at risk.

\subsection{Prediction of Future Failures From a Single Group}
\label{sec: binom}
For single-group prediction, we assume $i = 1 \dots n$ units are in service and observed until time $t_{ci} > 0$.  We define $t_{ci}$ in terms of operating time (e.g., days in service) rather than calendar time.  While we may observe units up until a data-freeze date, unless all units are placed into service at the same time, the amount of  operating time will vary among surviving units. A parametric lifetime model, $F(t;\bm{\theta})$, describes the lifetime distribution and by time $t_{ci}$ we have observed $r>0$ failures.  Thus, the remaining $n-r$ units have not failed (i.e, are right-censored). Let $Y$ be the number of additional failures that will occur between $t_{ci}$ and some future time $t_{wi}$, where $t_{wi} = t_{ci} + \Delta t$ ($\Delta t > 0$). Then, conditional on the $r$ failure times, $Y$ has a Binomial$(n-r, \rho)$ distribution, where
\begin{equation*}
\label{equation:single}
\rho= \frac{F(t_{wi};\bm{\theta}) - F(t_{ci};\bm{\theta})}{1-F(t_{ci};\bm{\theta})}.
\end{equation*} 

If $\rho$ were known, the lower and upper prediction bounds for the number of future failures could be obtained from the $\alpha$ and $1-\alpha$ quantiles of $Y$ 
$$\utilde{Y}_{\alpha} =\text{qbinom}(\alpha, n-r, \rho),\quad
\tilde{Y}_{1-\alpha} =\text{qbinom}(1-\alpha, n-r, \rho).$$

\noindent Otherwise, there are several methods to obtain these bounds described in the references in Section \ref{sec:intro}. In this paper, we use Bayesian prediction methods, which are discussed in more detail in Section \ref{sec:bayes}.

\subsection{Prediction of Future Failures across Multiple Groups}
\label{sec: poibinom}
Predicting the number of future within-sample failures across multiple groups is more challenging than prediction for a single population.  Using the same notations as above, we now add the subscript $g=1 \dots G$ to denote the heterogeneous groups or subpopulations. For a given group, $g$, with $n_g$ units, the probability of unit $i$ failing within the interval $(t_{c,ig},t_{w,ig})$  where $t_{w,ig} = t_{c,ig} + \Delta t$ is,
\begin{equation*}
\label{equation:rho}
\rho_{ig}= \frac{F(t_{w,ig};\bm{\theta_g}) - F(t_{c,ig};\bm{\theta_g})}{1-F(t_{c,ig};\bm{\theta_g})} \quad i=1 \dots n_g.
\end{equation*} 
The total number of failures in the population over $\Delta t$ is $Y = \sum_{g=1}^{G}\sum_{i=1}^{n_g}I(T_{ig}\in(t_{c,ig}, t_{w,ig}])$.  Conditional on the observed data and censoring times, the distribution of $Y$ is Poisson-binomial.  Similar to the single group case if the $\bm{\rho_g} = (\rho_{1g}, \dots \rho_{n_g g})$ values are known, the prediction interval endpoints are obtained from the $\alpha$ and $1-\alpha$ quantiles of $Y$ as follows, 

$$\utilde{Y}_{\alpha} =\text{qpoisbinom}(\alpha, \bm{\rho_{g}}),\quad
\tilde{Y}_{1-\alpha} =\text{qpoisbinom}(1-\alpha, \bm{\rho_{g}}).$$
Computing the cdf of the Poisson-binomial distribution is challenging. \cite{hong2013computing}, however, provides an algorithm to evaluate the Poisson-binomial distribution exactly or by an approximation. Because of the relatively small number of expected events (relative to the size of the risk set), we use the Poisson approximation in this paper.  The next section shows how to apply the single and multiple group prediction approaches within a Bayesian framework when the model parameters are not known.


\subsection{Bayesian Prediction Methods}
\label{sec:bayes}
Bayesian prediction methods require draws from the joint posterior distribution of the model parameters. The predictive distribution for the random variable $Y$ is  
\begin{equation}
 J(y) = \int_{\bm{\theta}} J(y;\bm{\theta})f(\bm{\theta}|t)d \bm{\theta}
\label{equation:predcdf}
\end{equation}
where $J(y; \bm{\theta})$ is the cdf of $Y$ and $f(\bm{\theta}|t)$ is the joint posterior for the model parameters.  Given a set of posterior draws, $\bm{\theta^{*}_j}$, $j=1 \dots B$, (\ref{equation:predcdf}) can be approximated by 
\begin{equation}
J(y) \approx \frac{1}{B}\sum_{j=1}^{B} J(y;\bm{\theta^{*}_j}).
\label{equation:estcdf}
\end{equation}

\noindent A $100(1-\alpha)\%$ equal-sided Bayesian prediction interval is given by the $\alpha/2$ and $1-\alpha/2$ quantiles of $J(y)$. A point prediction for the number of failures can be defined by the median (or the mean) of the predictive distribution to estimate  $\text{E}({Y})$.

In cases where it is difficult to evaluate $J(y;\bm{\theta})$ directly, another option is to simulate new values $Y^{*}_j$ from $J(y;\bm{\theta^{*}_j})$ for $j=1 \dots B$ so that $J(y) \approx (1/B)\sum_{i=1}^B I(Y^{*}_j \leq y)$. One limitation of this method is that it requires an extra layer of simulation that can be computationally demanding because the number of draws required to control Monte Carlo error, and provide the same precision as computing $J(y; \bm{\theta})$ directly, can be considerably larger.  In our applications, however, because the predictand is discrete, we found this additional simulation step did not require extra draws to obtain the same level of precision. That is, when we compared the two methods with $B=1{,}500$ in repeated trials  the methods usually gave the same interval endpoints. In the exceptional cases, the small proportion of off-by-one deviations was similar for the two methods.

\section{Application to the Motivating Examples}
\label{sec:application}
\subsection{Heat Exchanger Tubes}
\label{sec:heatexample}
\begin{figure}[t!]
\begin{tabular}{cc}
(a) & (b) \\[-4ex]
\splusbookfiguresizexx{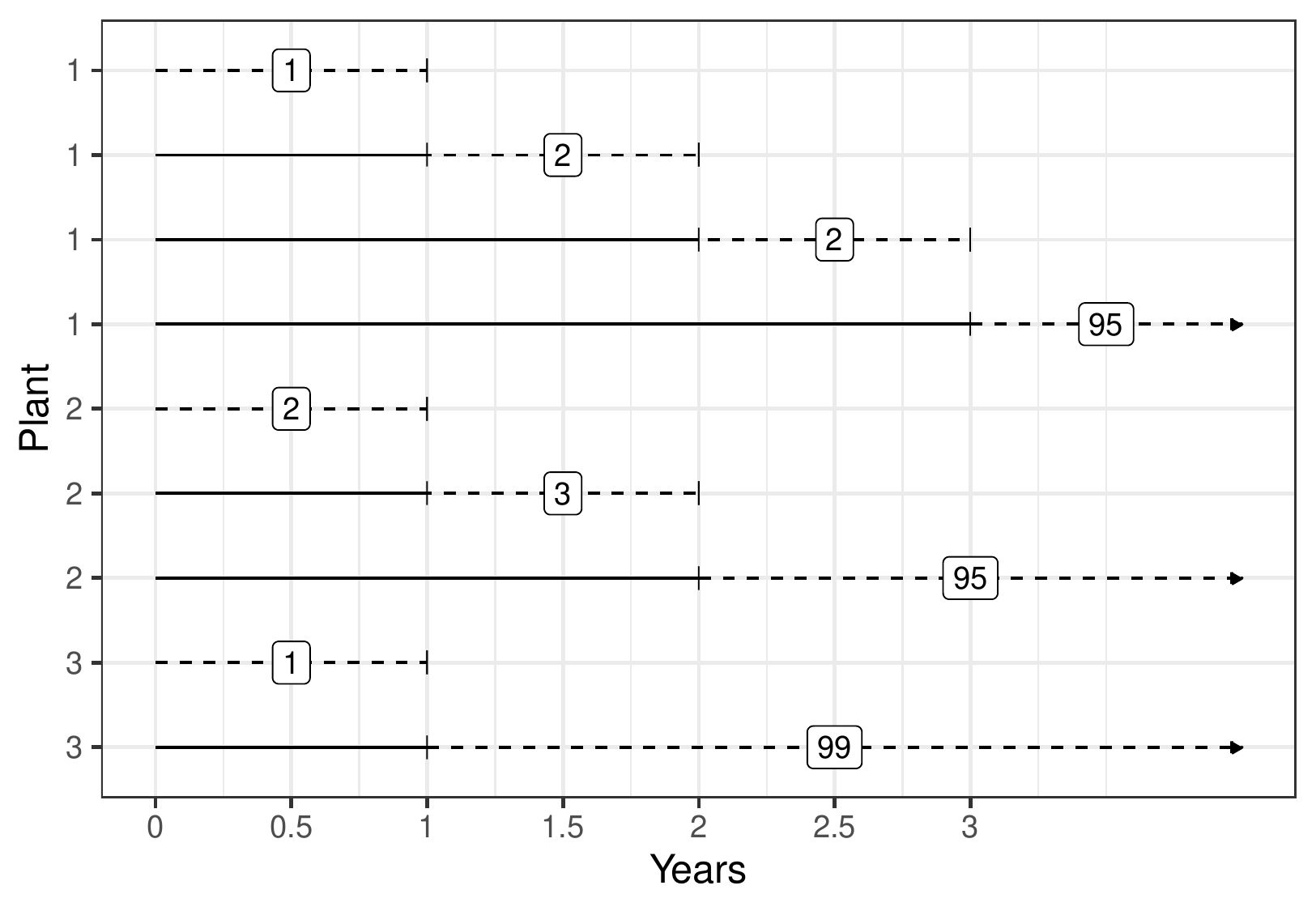}{3in}&
\splusbookfiguresizexx{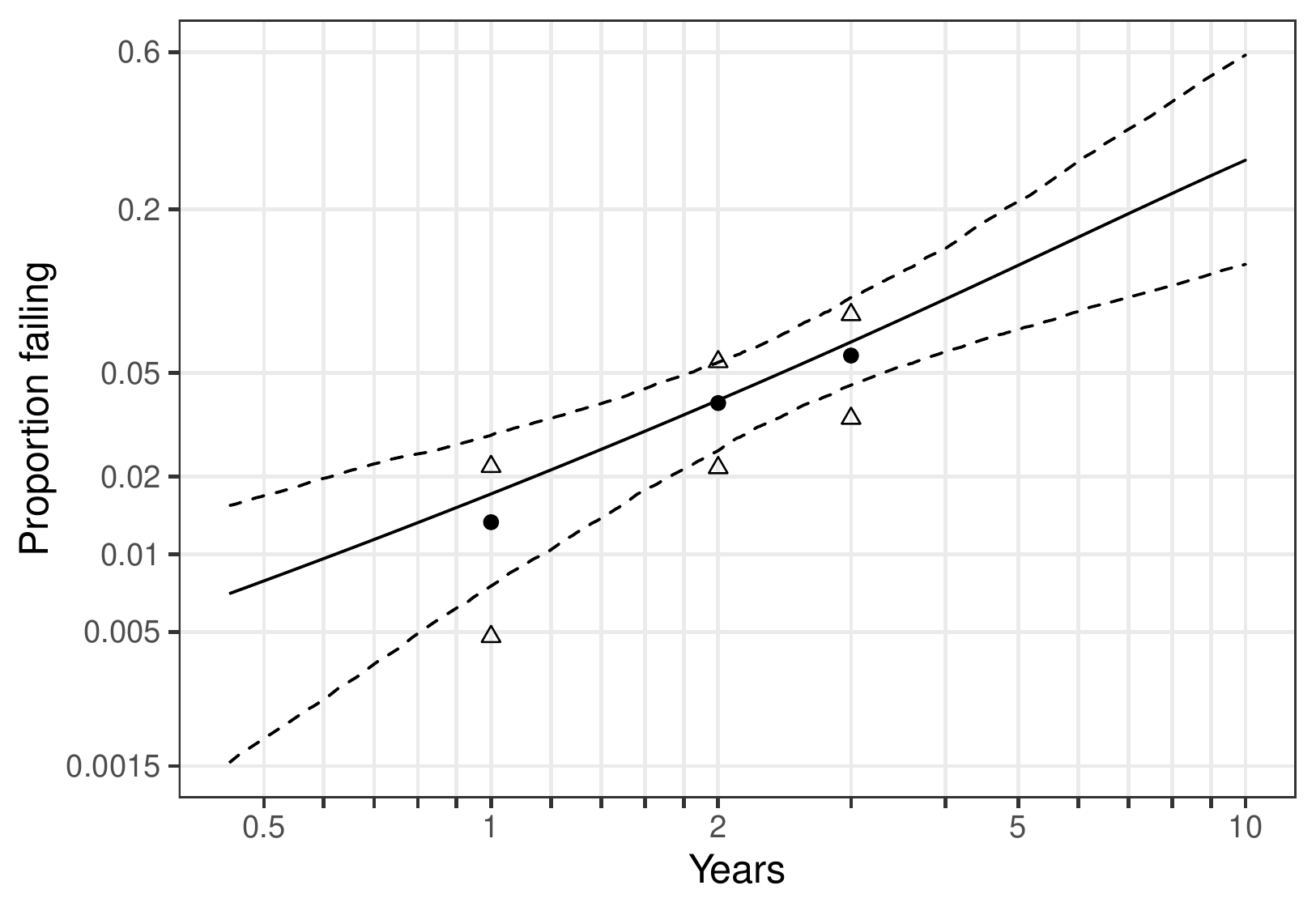}{3in}
\end{tabular}
\caption{(a) \footnotesize Event plot for heat exchanger tube inspection data in operating time.  Dashed arrows indicate right-censored tubes.  Dashed line segments indicate left- or interval-censored tubes. The number indicates the count. (b) The estimated pooled Weibull model (with weakly informative priors) plotted on Weibull probability scales. The solid curve corresponds to the median of the posterior draws as a function of plant age; pointwise 90\% credible bounds are indicated by the dashed lines. The solid circles are 
nonparametric estimates and the triangles are 90\% pointwise confidence intervals. }
\label{f:post_plant1}
\end{figure}

We extend the analysis of \cite{nordman2002weibull} by using a Bayesian hierarchical model and estimating the Weibull shape parameter (instead of assuming it is given). Then we predict the number of future failures using the methods outlined in Section \ref{sec:predfail}. In the case of the heat exchanger data, we have 3 groups ($G=3$) corresponding to 3 nuclear power plants with $n=300$ tubes, of which 11 have failed (4 failures are left-censored and 7 are interval-censored) at $t_c = 3$ years of service for the oldest plant, as illustrated in Figure \ref{f:post_plant1}(a).

To assess whether the Weibull distribution is appropriate, we combined the data from the three plants and estimated a pooled Weibull model.  Figure \ref{f:post_plant1}(b) plots the posterior median of the pooled Weibull model with axes on the Weibull probability scale. The plot also contains 90\% pointwise credible bands and a nonparametric estimate.  The Weibull distribution describes the data quite well, and the credible bands are consistent with the nonparametric point estimates.

For the hierarchical parameters, we specify the following proper prior distributions. Following the approach used in \citet[Section 15.2.2]{intervals} we will use diagonal braces ($<,>$) to refer to $95 \%$ central probability intervals, rather than the standard model parameters when specifying prior distributions for non-normal distributions to ease interpretability (such intervals are easier to elicit or otherwise specify).  We specify two different priors for $\eta_\sigma$, corresponding to weak ($\eta_{\sigma, \text{weak}}$) versus informative prior ($\eta_{\sigma, \text{inform}}$) information about the Weibull shape parameter. Weakly informative priors are constructed to be diffuse, relative to the likelihood and known scale of the data, and are designed to put probability mass on reasonable values of the parameters while down weighting nonsensical values that would not be consistent with what would be expected in the data.  In hierarchical models with many parameters and small amounts of data, weakly informative priors are especially useful as they can produce more stable estimates than maximum likelihood estimation or a Bayesian model with improper noninformative priors \citep{gelman2017prior}. If the failure is due to a wearout mechanism, then it is known that $\sigma < 1$.  Thus, for our informative prior, we put the majority of probability mass on values of $\sigma$ that are  less than 1.
\begin{equation*}\eta_{\sigma, \text{weak}} \sim \op{Lognormal} <0.08, \ 4.0 >,\
\eta_{\sigma, \text{inform}} \sim \op{Lognormal}< 0.37, \ 1.0 >,
\end{equation*}
\begin{equation*}
\eta_{t_p} \sim \op{Lognormal}<0.63,  \ 31.78>
\end{equation*}

\noindent We reparameterize the Weibull
distributions by replacing the usual scale parameter with the 0.05 quantile ($p=0.05$), which can be viewed as an alternative scale parameter. We do this because typically  we only have information about failures in the lower tail of the distribution and thus it would be easier (and more sensible) to set a prior distribution on such a parameter. For the standard deviation parameters ($\tau_{\sigma}, \tau_{t_{p}}$) we use half-$t$ distributions (with degrees of freedom $=4$), as suggested by \cite{gelman2006prior} for hierarchical standard deviation parameters when the number of groups is small.

\begin{figure}[t!]
\centering
  \includegraphics[width=.9\textwidth]{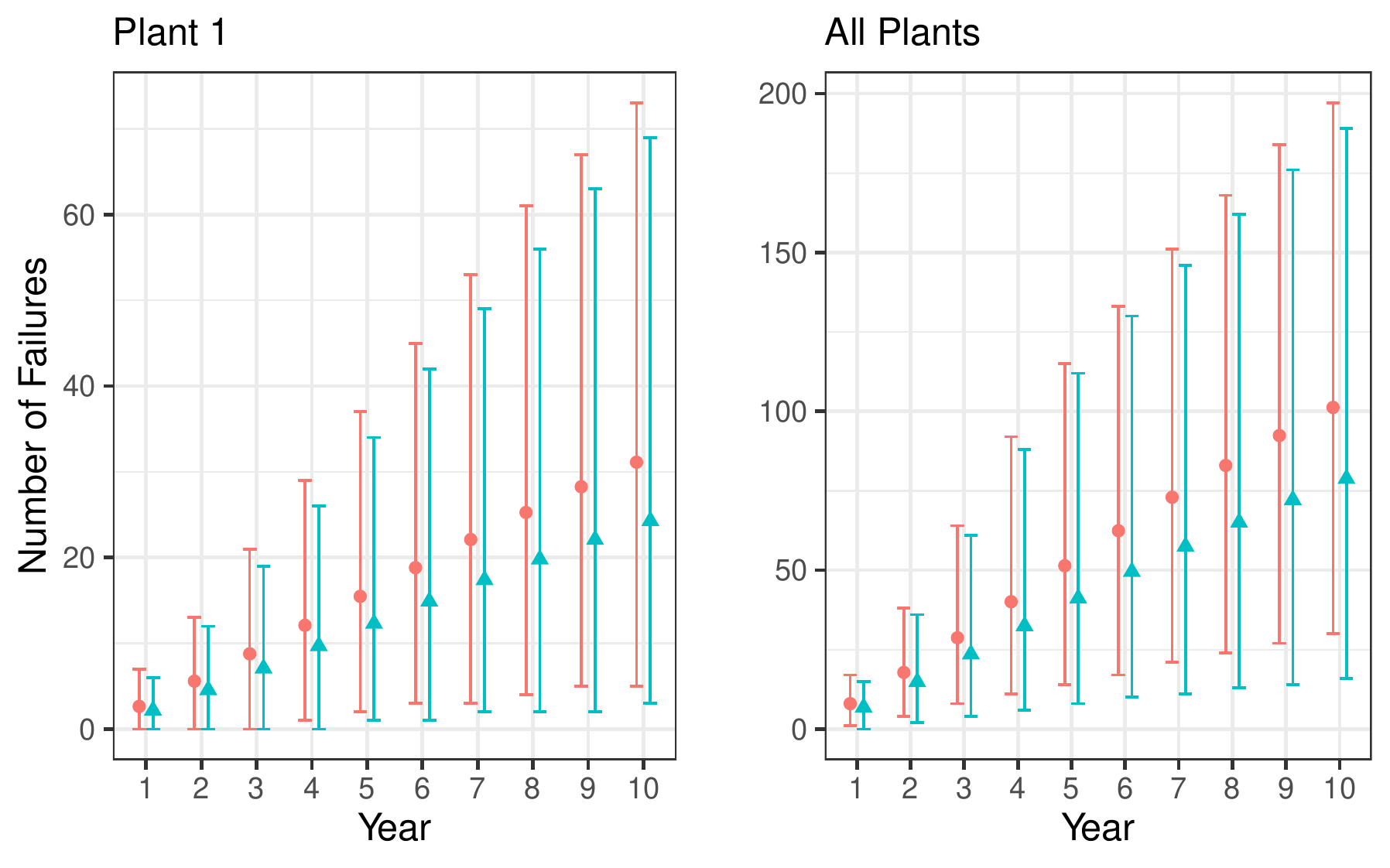}
  \caption{\footnotesize Yearly 95\% prediction intervals for the cumulative number of cracked heat exchanger tubes for Plant 1 (left), and all 3 plants (right).  Circles correspond to the model with an informative prior distribution on the shape parameter.  Triangles corresponds to the model with a weakly informative prior distribution on the shape parameter.}
  \label{pred-heat}
\end{figure}

Figure \ref{pred-heat} provides 95\% Bayesian prediction intervals using the Poisson approximation to the Poisson-binomial distribution.  The widths of the intervals are roughly the same, but the more informative prior distribution on the Weibull shape parameter results in intervals that are more conservative (i.e., intervals that predict a larger number of failures).  This difference becomes larger as we predict further into the future, which makes sense as the hazard rate increases over time. When a tube is found to be cracked, it is filled and thus taken out of service. Heat exchangers typically have excess capacity to allow between 5\% and 10\% of their tubes to be removed from service before requiring the entire heat exchanger to be taken out of service \citep{nordman2002weibull}.  Therefore, we predict out to the first year when the lower bound of our prediction interval exceeds 30 tubes (10\%), which occurs at year ten for the model with the informative prior distribution.  The widths of the prediction intervals increase as we predict further into the future.  In practice, however, such prediction intervals would be updated over time as additional data become available from future inspections (e.g., annually).  This would result in narrower prediction intervals.

\subsection{Backblaze Disk Drives}
As of the first quarter of 2019, Backblaze was collecting and reporting data on 85 different drive-models.  Some drive-models have been running since 2013 or before, while others were added at a later date. We use data through the first quarter of 2016 to fit the GLFP model, and use data starting the second quarter of 2016 through the first quarter of 2019 for validation. Because Backblaze periodically retires old drive-models and replaces them with newer, larger capacity models, only 28 out of the 44 drive-models are in both the testing and prediction data sets. A minimum of three failures was the criterion for inclusion of a
drive-model into our test data. Figure 3(a) shows a scatterplot of the number of active drive-models (i.e., drives in the risk set) at our data-freeze date, March 31, 2016.

To complete the GLFP model, we specify the following proper prior distributions.  We consider these prior distributions weakly informative because they put probability mass on a wide range of values for all model parameters---much larger than the ranges that would be expected from typical applications where Weibull distributions are used to describe  reliability. We use
\begin{equation*}
\label{eq:hier-model}
\sigma_{1} \sim  \op{Lognormal} <0.14, \  7.1 >, \quad t_{p_{1}} = \exp\left(\mu_{1} + \sigma_{1}\,\Phi_\textrm{{sev}}^{-1}(p_1)\right)  \sim \op{Lognormal} < 22,  \ 5.5 \times 10^4 >,
\end{equation*}
\begin{equation*}
\eta_{\pi} \sim \op{Normal}(-3, 1^2), \quad \eta_{\sigma_2}  \sim \op{Normal}(0, 2^2), \quad \eta_{t_{p_2}}  \sim \op{Normal}(9, 2^2).
\end{equation*}

\noindent We reparameterize the Weibull
distributions by replacing the usual scale parameter with the 0.50 quantile for the early failure mode
($p_1=0.50$) and the 0.20 quantile for the wearout failure mode
($p_2=0.20$). For the hierarchical standard deviation parameters, we use half-Cauchy prior distributions because we have a larger number of groups and observed failures compared to the heat exchanger example.  For more details on the choice of these prior distributions, see \cite{mittman}.  

\begin{figure}[t!]
\begin{tabular}{cc}
(a) & (b) \\[-.5ex]
\splusbookfiguresizexx{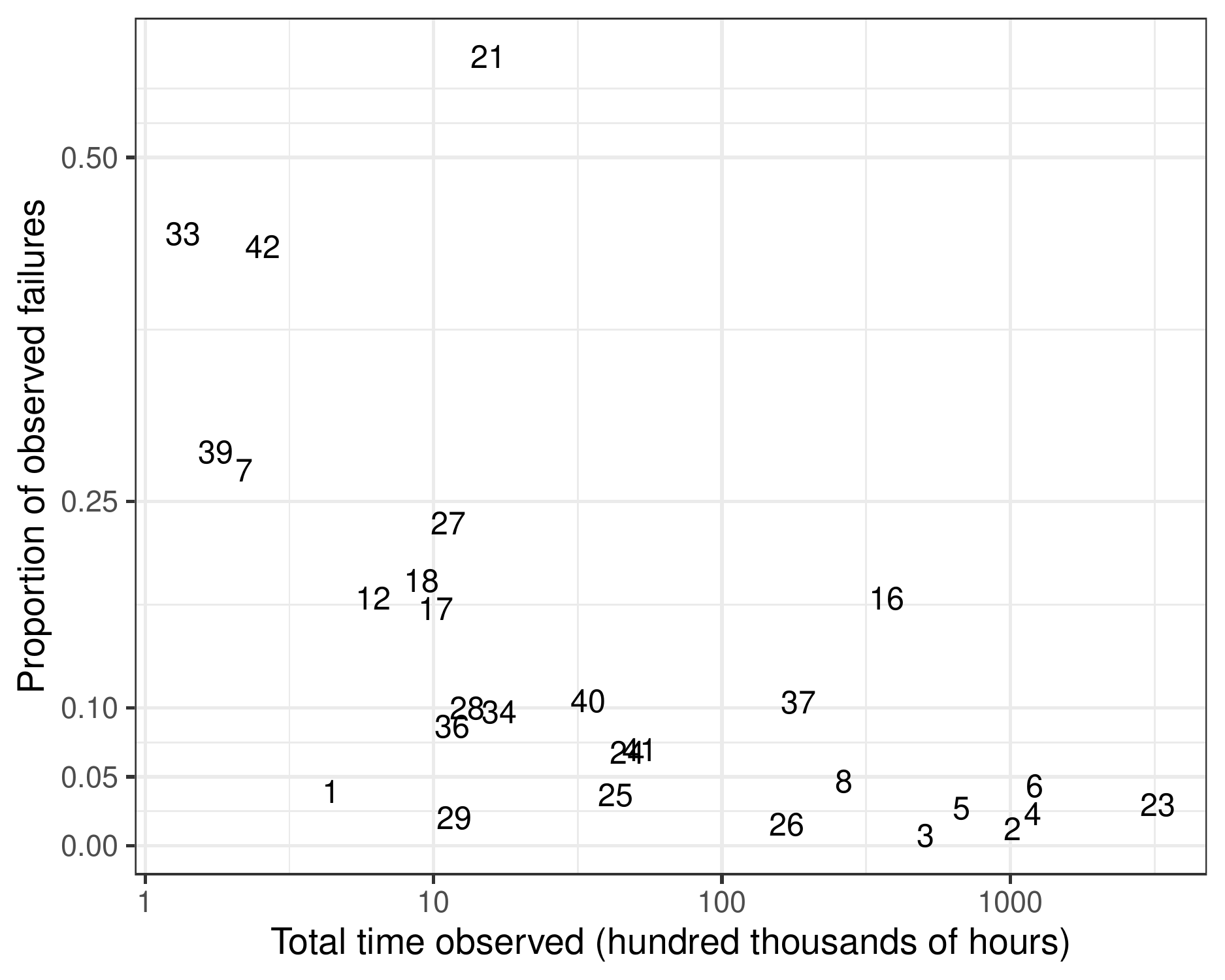}{3in}&
\splusbookfiguresizexx{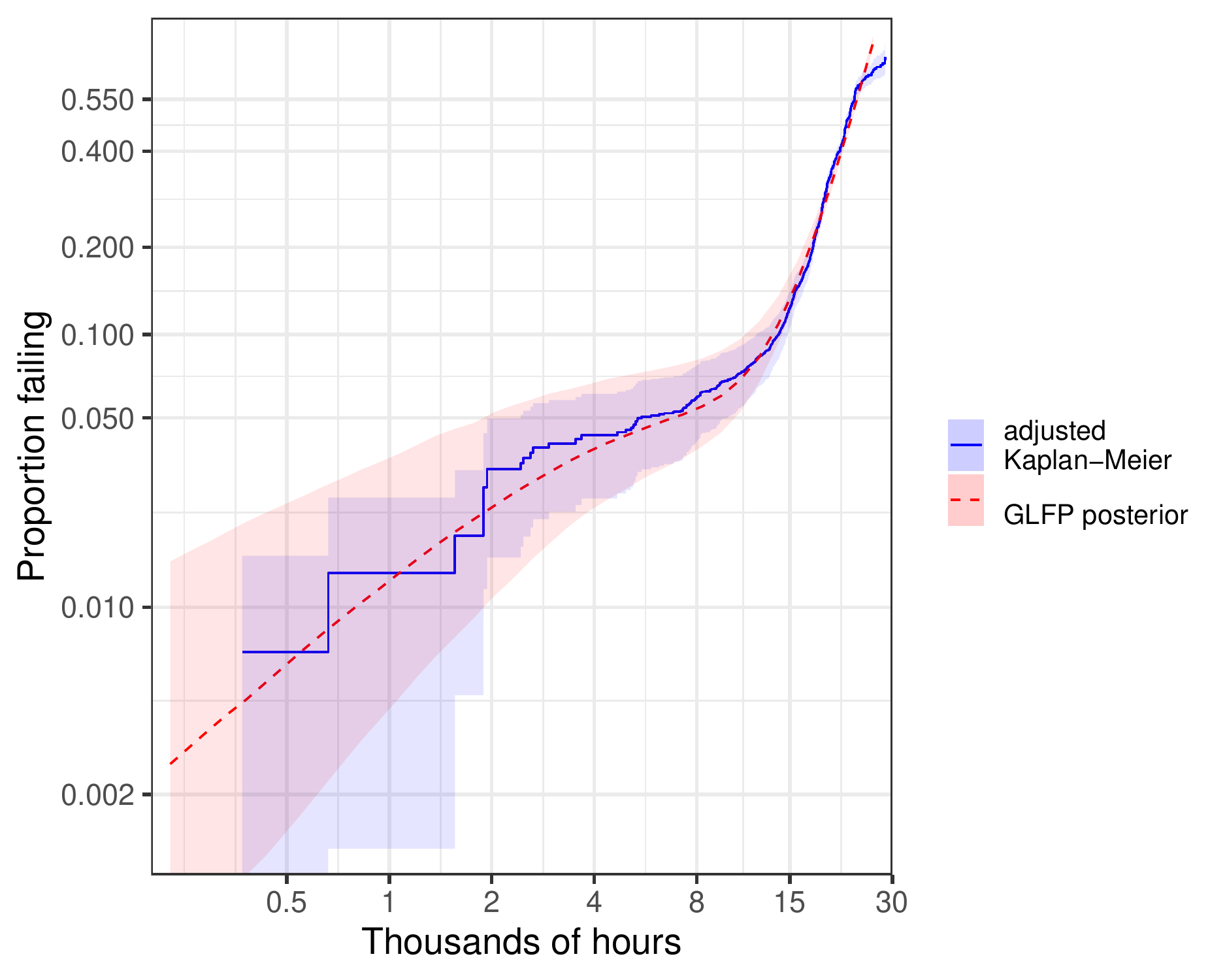}{3in}
\end{tabular}
\caption{(a) \footnotesize Scatterplot of the total observed running time (hundred thousands of hours) versus the proportion of observed failures. Each number indicates a unique drive-model (see supplemental material for corresponding brand name). The $y$-axis is linear. The $x$-axis is on a log scale. ~(b) The estimated GLFP model for Drive-model 14 plotted on Weibull probability scales.
The dashed line corresponds to the median of the posterior draws; pointwise 90\%
credible bounds are indicated by the smooth region. The solid step function is an adjusted
Kaplan-Meier estimate with 90\% pointwise confidence intervals. }
\label{f:bbdata}
\end{figure}

Figure \ref{f:bbdata}(b) shows an overlay of the posterior median of the fitted GLFP model for Drive-Model 14 (the drive-model with the most observed failures) onto an adjusted Kaplan-Meier estimate with axes on the Weibull probability scale. The plot also contains 90\% pointwise credible intervals for both estimates. The GLFP model fits well, as it is able to describe the rapid increase in the empirical cdf between 8,000 and 20,000 hours.

\subsubsection{Dynamic Risk Set}
\label{s:riskset}
When making within-sample predictions for a large population or fleet of systems there is often a need to adjust to the risk set over time.  In addition to units failing, units could be removed temporarily from the risk set for inspection or maintenance.  Due to advancements in technology, subgroups of products are also frequently retired as they are replaced by newer technology.  Also, in some applications units that will be added to the population in the future may need to be included at the appropriate time.  Accurately adjusting the risk-set to reflect the changing sample size in the prediction population is important to avoid biased predictions for the number of future failures.  

\subsubsection{Disk Drive Failure Predictions}
\label{s:bbpreds}
\begin{figure}[t!]
  \includegraphics[width=.9\textwidth]{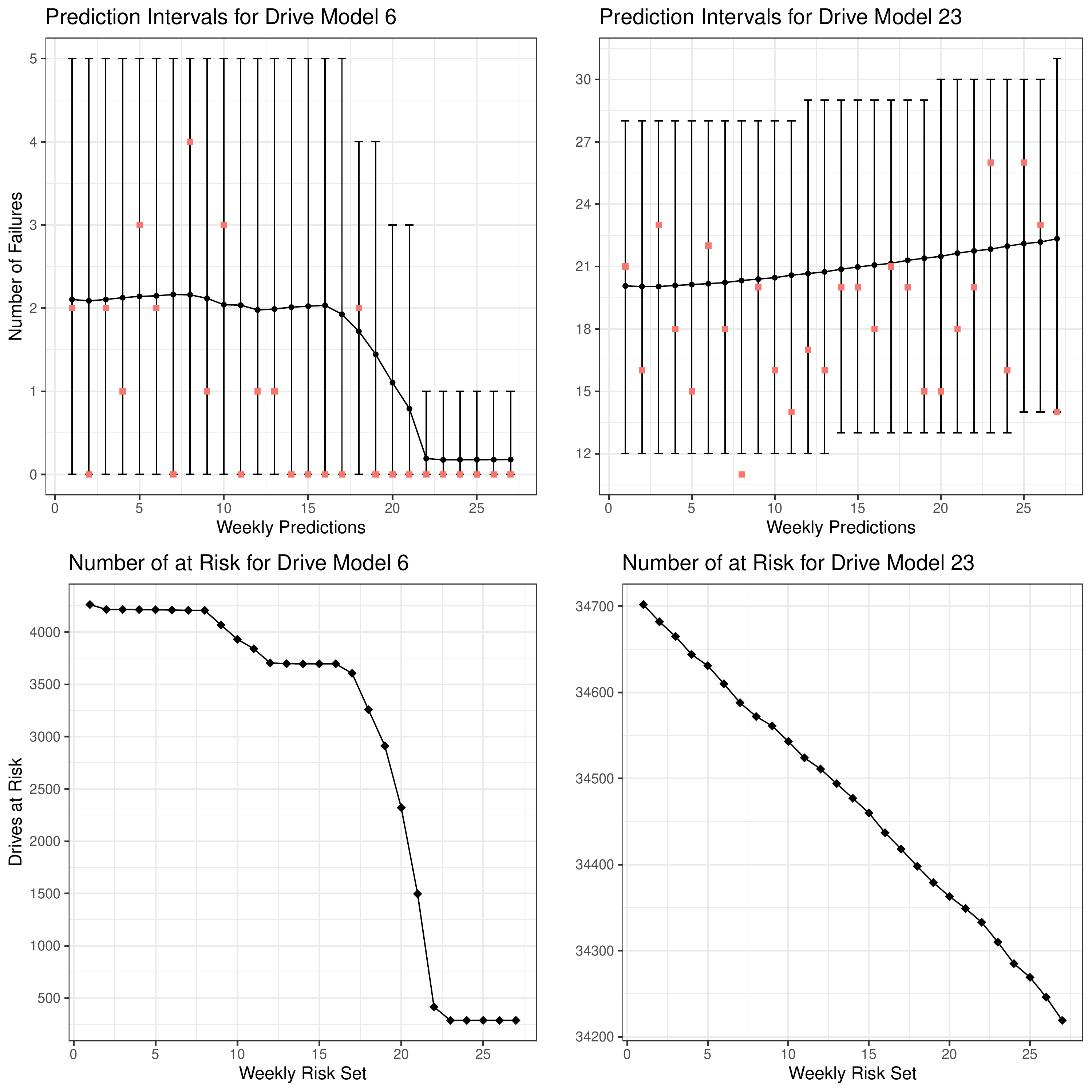}
  \caption{\footnotesize Top: weekly 95\% prediction intervals and predicted average number of failures (dots) for Drive-Models 6 and 23.  Squares are the observed number of failures.  Bottom: weekly risk set (diamonds) for Drive-Models 6 and 23.}
  \label{f:pred_fails}
\end{figure}

As mentioned in Section \ref{sec:bbdata}, Backblaze periodically retires certain subpopulations of drive-models from the population before they fail, and replaces them with newer more efficient disk drives that have larger storage capacity.  Backblaze does not disclose their decision-making policy for retiring old technology, but their website says they regularly migrate older drive-models out for newer more efficient drives. Their data reflect such actions. The bottom row in Figure \ref{f:pred_fails} plots the changing risk set size for Drive-Models 6 and 23 starting on the data-freeze date (Q2 2016).  For Drive-Model 6 (bottom left), at week 1 there are over 4,000 active drives in the risk set, but by week 26 there are fewer than 500 drives at risk of failing.  In contrast, Drive-Model 23 (bottom right) has drives leaving the within-sample prediction population at a roughly linear rate.  To avoid making predictions for drives that are no longer in the sample (either due to failure or removal), we predict using one-week intervals (i.e, $\Delta t = 1 \text{ week}$) for 26 weeks. Each week we update the risk set by removing drives from the prediction sample that are no longer in service (either because they failed or were taken out of service).  We make weekly predictions for the next 26 weeks after the data-freeze date.  At the end of 26 weeks, we stop predicting because the majority of the drive-models have few units remaining in the risk population.

The top two plots in Figure \ref{f:pred_fails} show the 95\% weekly prediction intervals for the number of future failures for Drive-models 6 and 23.  The dots are the median of the predictive distribution (a point prediction) and the squares are the observed number of failures.  The prediction intervals for both drive-models cover the observed number of failures (see the supplementary material for the plots of the prediction intervals for all drive-models).  The gradual decrease in the number of predicted failures for Drive-Model 6 (top-left) is because Backblaze started retiring units from that model between weeks 15 and 20 of the prediction period.  In contrast, fewer than 100 units were retired from Drive-Model 23, which explains the monotonic increase in the predicted number of failures from week 1 to 26.

\subsection{Product A Returns}
\label{s:prodAreturn}
\subsubsection{Distribution Choice}
The Product A data exhibits different subpopulations with seasonal behavior---a feature discovered by \cite{shan2020seasonal} during their exploratory data analysis.  To create groups that are more homogeneous, they used the empirical monthly repair rate (the number of returns divided by the number of active contracts) for each state and province and applied the bottom-up/agglomerative hierarchical clustering to identify four clusters having similar seasonal behaviors. We use these clusters for our modeling of the Product A data. 
\begin{table}[t!] 
\begin{center}
\begin{adjustbox}{width=1\textwidth}
\begin{tabular}{@{\extracolsep{0pt}} crrrrrrrrr} 
\toprule
 &\multicolumn{2}{c}{Cluster 1}
&
\multicolumn{2}{c}{Cluster 2} 
&
\multicolumn{2}{c}{Cluster 3} 
&
\multicolumn{2}{c}{Cluster 4} \\
\hline
Entry-Month  &\thead{Number of \\ Contracts} & \thead{Number of \\ Returns}    & \thead{Number of \\ Contracts} & \thead{Number of \\ Returns}  & \thead{Number of \\ Contracts} & \thead{Number of \\ Returns} & \thead{Number of \\ Contracts} & \thead{Number of \\ Returns}     \\
\bottomrule

 January & 1,441 & 140 & 904 & 67 & 527 & 48 & 704 & 69 \\ 
 February & 2,050 & 233 & 1,182 & 113 & 644 & 64 & 954 & 93 \\ 
 March & 2,119 & 217 & 1,514 & 153 & 678 & 54 & 1,088 & 78 \\ 
 April & 1,906 & 185 & 1,192 & 106 & 661 & 62 & 848 & 70 \\ 
 May & 1,768 & 175 & 1,342 & 119 & 740 & 54 & 1,109 & 82 \\ 
 June & 2,543 & 235 & 1,736 & 154 & 1,232 & 108 & 1,729 & 143 \\ 
 July & 2,097 & 128 & 879 & 60 & 758 & 55 & 1,302 & 63 \\ 
 August & 2,648 & 184 & 1,115 & 99 & 851 & 60 & 1,817 & 89 \\ 
 September & 3,301 & 296 & 2,140 & 163 & 1,042 & 82 & 1,374 & 114 \\ 
 October & 3,974 & 337 & 3,230 & 209 & 966 & 98 & 1,232 & 126 \\ 
 November & 3,472 & 259 & 2,648 & 164 & 744 & 70 & 927 & 74 \\ 
 December & 2,303 & 193 & 1,610 & 100 & 684 & 46 & 788 & 69 \\ 
 \hline
 Total & 29,622 & 2,582 & 19,492 & 1,507 & 9,527 & 801 & 13,872 & 1,070 \\ 
\hline
\end{tabular} 
 \end{adjustbox}
  \caption{\footnotesize Number of units entering service (and number of returns) by entry-month for each cluster.} 
  \label{table: prodA} 
  \end{center}
\vspace{-1cm}
\end{table}

Table \ref{table: prodA} shows the number of units entering service by entry-month, as well as the number that had a warranty return. We will use Product A data from 2011 to 2016 for model building, incorporating time-varying as well as constant covariates, and assess predictive performance using monthly predictions for the number of returns in each month of 2017.

\begin{figure}[t!]
\centering
  \includegraphics[width=.9\textwidth]{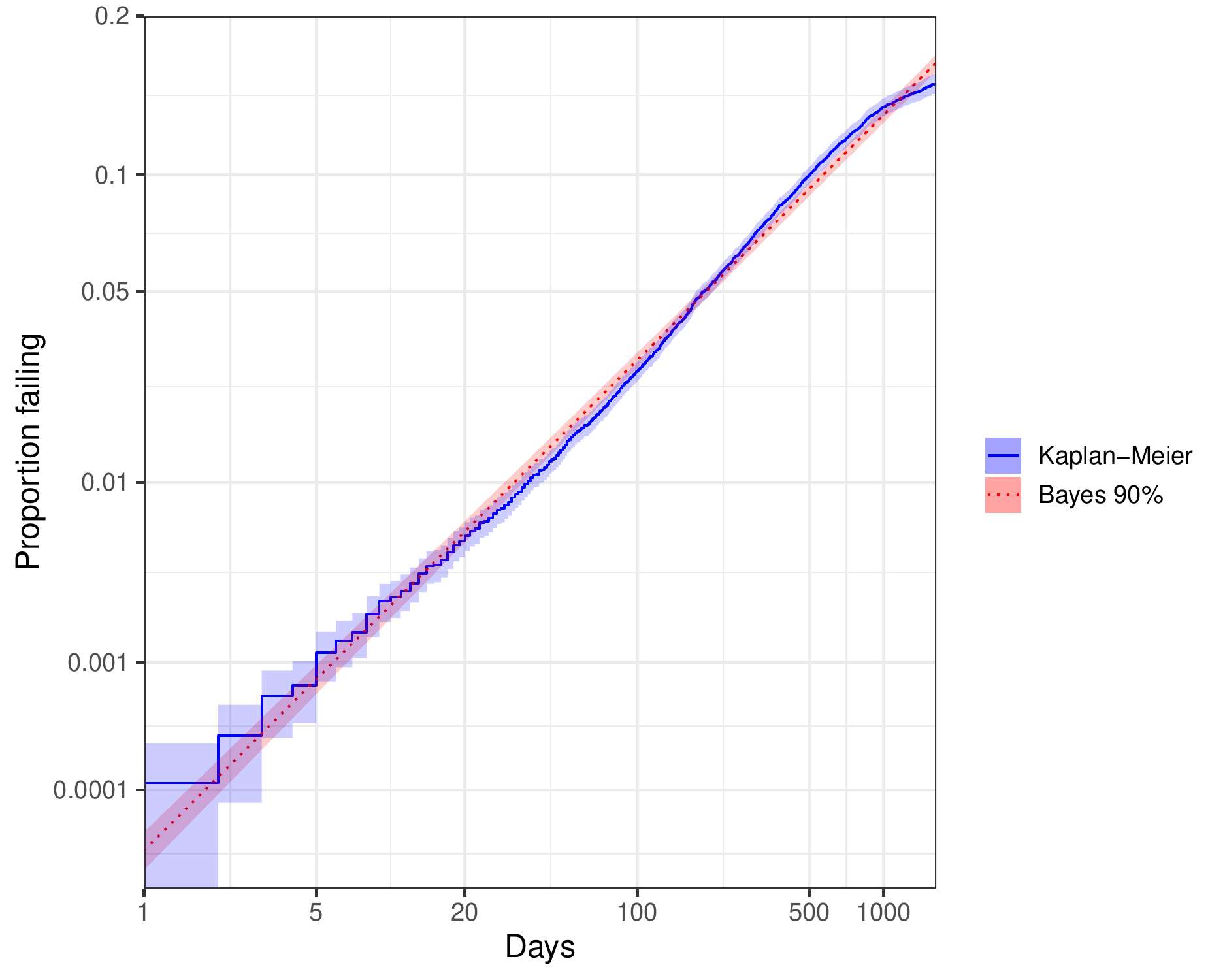}
\caption{\footnotesize Estimated  Fr\'echet distribution for Product A time-to-return data from Cluster 1 in the U.S., plotted on Fr\'echet probability scales.  The dashed line corresponds to the median of the posterior draws; pointwise 90\% credible intervals are indicated by the smooth region.  The solid step function is a Kaplan-Meier estimate with 90\% pointwise confidence intervals.}
  \label{f:prodA}
\end{figure}

Before we incorporate covariate information into our model, we require a lifetime distribution that adequately approximates the observed data.  We select the Fr\'echet distribution,  another log-location-scale distribution because it provides (unlike the Weibull and lognormal distributions) a good description of the Produce A time-to-first-return data. Using weakly informative priors, we fit the Fr\'echet distribution to Product A systems from the United States (this subpopulation has the most observed returns).  Figure \ref{f:prodA} plots the posterior median of the Fr\'echet distribution as well as a  Kaplan-Meier estimate.  The uncertainty intervals between the Fr\'echet and nonparametric estimates overlap well, indicating the Fr\'echet is an appropriate distribution. Both sets of intervals are narrow because of the large number of returns.

\subsubsection{The Cumulative Damage Model}
We first use a cumulative damage (CD) (or cumulative exposure) model to take into account fixed and seasonal covariates following the general approach used in \cite{hong2013field}.  The CD model assumes that a system deterministically accumulates damage over time (e.g., as a function of days in service) until it reaches a latent random threshold, at which time it is returned.  By incorporating covariates, systems can accumulate damage more rapidly in certain months or locations when compared to other months or locations. For unit $i$, $x_i(d)$ is the value of the covariates on day $d=0 \dots d_{n_i}$ where $d_{n_i}$ is the total number of days the unit was in service before the data-freeze date.  Let $\bm{x}_i(d)$ be the observed covariate history from day 0 to day $d$. Then the deterministic latent cumulative damage for unit $i$ in cluster $c$ by $d_{n_i}$, which is its final day of observation, is 
\begin{equation*}
u_i = u[d_{n_i};d, \alpha_c, \bm{\beta_c},\bm{x}_i(d_{n_i})] = \int_{0}^{d_{n_i}} \text{exp}[\zeta(s)]ds = \sum_{s=0}^{d_{n_i}} \text{exp}[\zeta(s)],
\end{equation*}
where $\zeta(s)$ is the amount of damage on day $s$. Each unit has a random threshold $U_i$ with cdf $F_0(u) = \text{Pr}(U \leq u)$ and unit $i$ is returned when $u_i$ exceeds $U_i$. The cumulative damage is latent because the parameters in $\zeta(s)$ are unknown.

For the Product A, let $i=1\dots n = 63{,}132$ denote the number of systems in the warranty database.  Define $m=1 \dots   12$ the calendar month, and $c=1 \dots  4$ the cluster.  We let 
\begin{equation}
\zeta_{i,c}(s) = \alpha_c \text{Canada}_{i,c} + \sum_{m=1}^{12}\sum_{c=1}^{4}\beta_{c,m}\text{I}_{i,c,m}(s)
\end{equation}
where $\text{Canada}_{i,c}=1$ if unit $i$ from cluster $c$ is from Canada and $0$ otherwise, and the time-varying dummy variables are 

\noindent

$$
\text{I}_{i,c,m}(s)=
\begin{cases}
1 \text{ if on day } s \text{ unit } i \text{ from cluster } c \text{ is in calender month } m\\
0 \text{ otherwise} 
\end{cases} 
$$
Note that for unit $i$, the double sum on the right-hand side of $(7)$ is equal to $\beta_{c,m}$ when unit $i$ is in cluster $c$ and day $s$ is in month $m$. The $\bm{\beta_c} = (\beta_{c,1} \dots \beta_{c,12}) '$ parameters describe the effect on the amount of cumulative damage for cluster $c$ and month $m$.  The parameter $\alpha_c$ is a country fixed effect capturing if a unit in cluster $c$ is in service in the U.S. or Canada. Operationally, we define the matrix $\bm{Z}$ containing $z_{i,j} = $ number of days in service for unit $i$ in month $j$.  Also, we add an initial column  to $\bm{Z}$ for the Canada dummy variable. Then $u_i = u[d_{n_i};d, \alpha_c, \bm{\beta_c},\bm{x}_i(d_{n_i})]= \text{exp}([\alpha_c \ \bm{\beta_c}])\bm{z_i'}$. 

We use a Fr\'echet distribution ($F_0(t; \mu_0, \sigma_0$)) with parameter $\mu_0$, and parameter $\sigma_0$, for the cumulative damage threshold.
For a given cluster $c$, the likelihood is thus,
\begin{align*}
&L(\alpha_c, \bm{\beta_c}, \mu_0, \sigma_0|\text{Failure-time Data, Covariate History}) \\
 &=\prod_{i=1}^{n}\left(\zeta(d_{n_i})f_0(\text{exp}([\alpha_c \ \bm{\beta_c}])\bm{z_i^{'}};\mu_0, \sigma_0)\right)^{\delta_i} \times \left(1-F_0(\text{exp}([\alpha_c \ \bm{\beta_c}])\bm{z_i^{'}};\mu_0, \sigma_0)\right)^{1-\delta_i}
\end{align*}
where $\delta_i=1$ if a unit fails and $0$ if it is right-censored. Because the scale of the latent $u_i$ values is arbitrary, we set $\mu_0 =0$ to avoid over-parameterization.  For each of the four clusters, there are at least 46 failures within each month (cf., Table 1), which provides enough information to be able to fit a marginal cumulative damage model to each cluster.  For each cluster, we specify the following weakly informative prior distributions,
\begin{equation*}
\beta_{c,m}, \alpha_c \ind \op{Normal}(0, 10^2) \quad
\sigma_{c,0} \ind \op{Lognormal} < 0.02, \ 51.2 >.
\end{equation*}  
These prior distributions put probability mass on a wide range of values for all model parameters--much larger than we would expect given the data.  A sensitivity analysis of the prior distributions (e.g., doubling the standard deviations of the priors) did not affect the posterior distribution of the parameters. 
\subsubsection{The Proportional Hazards Model}
An alternative approach for incorporating time-varying covariates is to use a parametric proportional hazards (PH) model \citep{cox1972regression}.  In contrast to the CD model, where a unit's hazard depends on the entire covariate history of the unit, under the PH model, the hazard on day $d$ depends only on the covariate level on day $d$.  Denote the baseline hazard function by $\lambda_0(d; \mu_0, \sigma_0)$.  We use the same definition of $\zeta(s)$ as in (7) to define the time-varying covariates, so on a given day, $d$, the hazard function for a unit is 
\begin{equation*}
h(d;\mu_0, \sigma_0, \alpha_c, \bm{\beta_c}) = \lambda_{0}(d;\mu_0, \sigma_0)\text{exp}[\zeta(d)]
\end{equation*}

\noindent and the cumulative hazard function is
\begin{equation*}
H(d;\mu_0, \sigma_0, \alpha_c, \bm{\beta_c},\bm{x_i}(d)) = \int_{0}^{d}\lambda_0(s; \mu_0, \sigma_0)\text{exp}[\zeta(s)] ds.
\end{equation*}

We again use the Fr\'echet distribution to define our parametric baseline hazard function, $\lambda_0(d; \mu_0, \sigma_0)$. For a given cluster $c$, the log-likelihood (e.g., \citealt[Section 6.5.1]{lawless}) is
\begin{align*}
&l(\alpha_c, \bm{\beta_c}, \mu_0, \sigma_0|\text{Failure-time Data, Covariate History}) \\
 &=\sum_{i=1}^{n}\delta_i \text{log}(h(d_{n_i};\mu_0, \sigma_0, \alpha_c, \bm{\beta_c})) - H(d_{n_i};d, \mu_0, \sigma_0, \alpha_c, \bm{\beta_c},\bm{x_i}(d_{n_i}))
\end{align*}
where $\delta_i=1$ if a unit is returned and $0$ if it is right-censored. We specify the same weakly informative prior distributions for the unknown parameters that were used for the CD model.

\subsubsection{Prediction Results}

The monthly predictions for the number of warranty returns on the 2017 hold-out data for Cluster 1 are shown in Figure \ref{pred-prodA} (see the supplementary material for the prediction intervals for all of the clusters).  Predictions using the CD model (triangles) and the PH model (circles) are presented.  For Cluster 1, prediction intervals from the CD model capture the true number of returns fifteen out of twenty-four times compared to nineteen times for the PH model.  The PH model also has narrower prediction intervals, on average.  This suggests that the PH model might be better suited to the Product A data.  In terms of the assumptions behind the CD and PH models, this means the history of the seasonality is less important than the specific month the system is in operation. Viewed another way, over several years of service the effect of seasonality tends to average-out in the CD model, but not with the PH model.

\begin{figure}[t!]
\centering
  \includegraphics[width=.9\textwidth]{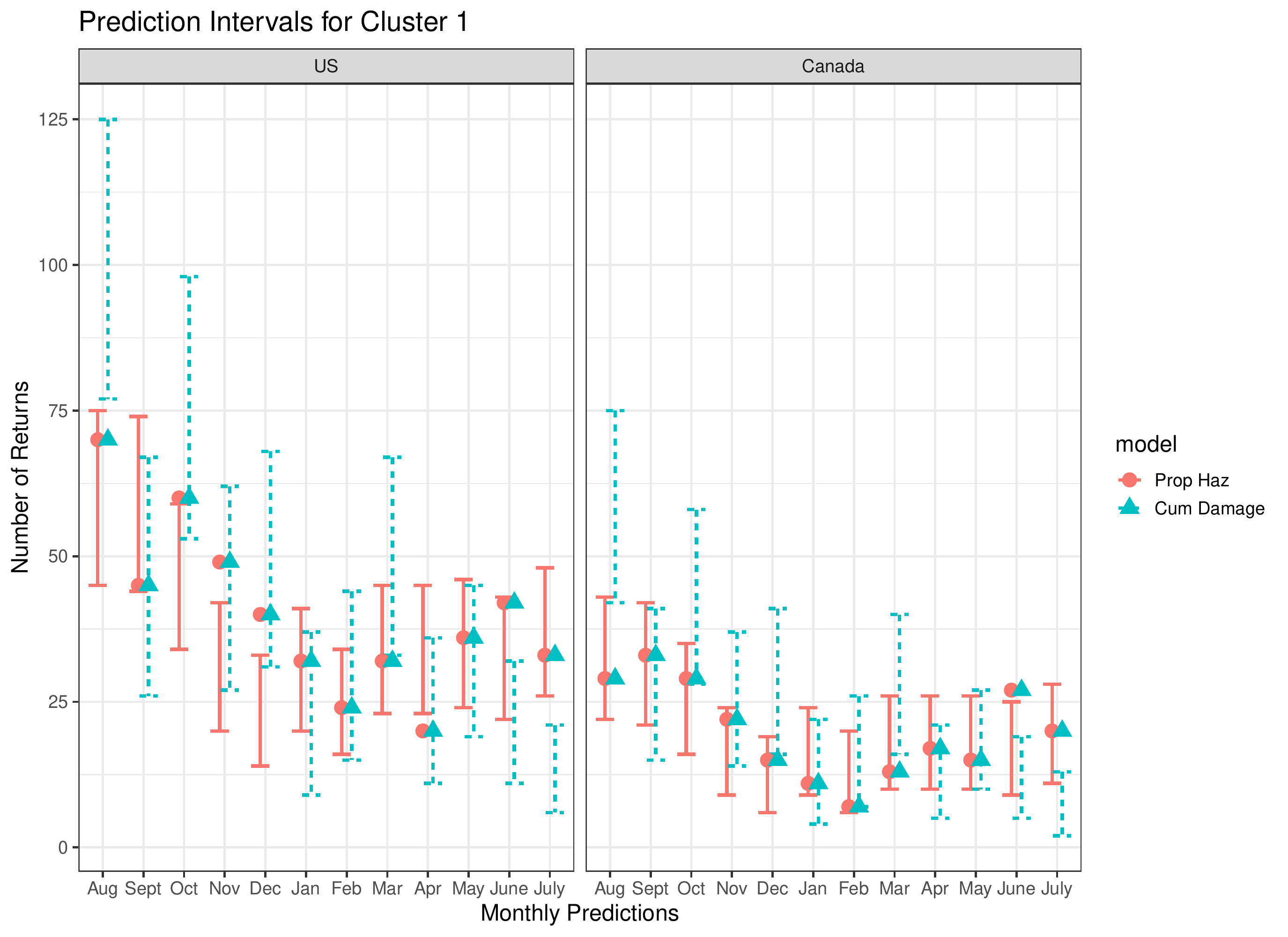}
  \caption{\footnotesize Monthly 95\% prediction intervals for the number of warranty returns for systems in Cluster 1 on the hold-out data.  The true numbers of returns are indicated by the circles and triangles.  The dashed lines (and triangles) are prediction intervals from the CD model; the solid lines (and circles) are the prediction intervals from PH model.}
  \label{pred-prodA}
\end{figure}

\section{The Simulation Experiment}
\label{sec:sim}
\subsection{Goals of the Simulation}
\label{sec:goals}
This section describes the design of our simulation study to examine the properties of Bayesian prediction intervals when applied to within-sample prediction for heterogeneous lifetime data with weakly informative prior distributions. For each simulation, we generate data, $t_{ig}$, from the following hierarchical model,
\begin{equation*}
T_{ig} \stackrel{ind}{\sim} \text{Weibull}\left(t_{p_g}, \sigma_{g} \right), \quad \sigma_{g} \ind \op{Lognormal} \left( \eta_{\sigma}, \tau^2_{\sigma} \right),
\end{equation*}
\begin{equation*}
\label{eq:hier-model}
t_{p_g} = \exp\left(\mu_{g} + \sigma_{g}\,\Phi_\textrm{{sev}}^{-1}(p)\right)  \ind \op{Lognormal} \left(\eta_{t_{p}}, \tau^2_{t_{p}}\right).
\end{equation*}

\subsection{Experimental Factors}
\label{sec:factors}
Our simulation has all units entering the population at the same time and they are observed until a fixed censoring time, $t_c$ (i.e., type I censoring).  The goal is to use the information from the observation period to predict the number of failures between ($t_c$, $t_w$] for  each subpopulation $(r_g)$, as well as the whole population.  This setup mimics certain real-world applications using product reliability data, where products enter service in a single cohort.
The particular factors in the simulation experiment used were
\begin{itemize}
\item $G$: the number of subpopulations
\item $t_{p_g}, \sigma_{g}$: the subpopulation specific random parameters
\item E$\big( r \big)$ = E$\Big( \sum_{g=1}^G r_g \Big)$: the expected number of failures before $t_c$
\item $p_f$: the expected proportion of failures before time $t_c$
\item $p_\delta$: the expected proportion of failures between times $t_c$ and $t_w$.
\end{itemize}
\subsection{Factor Levels}
In order to cover a range of situations, we used the following levels of the factors
\begin{itemize}
\item $G=5, 10$
\item E$\big( r \big)$ = 25, 50, 75, 100, 125 for $G=5$
\item E$\big( r \big)$ = 50, 100, 150, 200, 250 for $G=10$
\item $p_f$ = 0.01, 0.05, 0.10, 0.20
\item $p_\delta$ = 0.10, 0.20
\item $t_{p_g} \sim \text{Lognormal}< 4, \ 8 >$, $t_{p_g} \sim \text{Lognormal}< 10, \ 14 >$
\item $\sigma_{g} \sim \text{Lognormal}< 0.15, \ 0.40 >$, $\sigma_{g} \sim \text{Lognormal}< 0.50, \ 0.75 >$.  
\end{itemize}
We used different levels of E$\big( r \big)$  so the expected number of failures per group is the same for $G=5$ and $G=10$. For each factor-level combination, we simulated 300 data sets.  The number of units per group ($n_g$) is determined once the factor levels are set, $n_g = E\big( r \big)/(G \times p_f$).   To compute the censoring times for each factor-level combination, we simulated 50,000 sets of Weibull parameters from the hierarchical distributions.  We then averaged the Weibull cdfs across all the simulated parameters and numerically calculated $t_c$ and $t_w$ to obtain the correct $p_f$ and $p_\delta$ for each simulated data set.
\subsection{Estimation} 
We again use Bayesian estimation for our simulation study.  Performance of our prediction intervals is evaluated in terms of coverage probability. Here coverage probability is interpreted as the proportion of times that prediction intervals (for single and multiple groups) cover the true number of future within-sample failures.  We do not want our prior distributions to bias the results: diffuse prior distributions (e.g., a flat prior) put probability mass on extreme values not consistent with the range of practical situations we are trying to mimic; priors that are too informative can also affect the posterior by overly constraining the likelihood.  Therefore, for all factor level combinations, we specify the following weakly informative prior distributions, 
$$\eta_{\sigma} \sim \op{Lognormal} < 0.08, \ 4.0 >,\quad
\eta_{t_p} \sim \op{Lognormal}< 2.75, \ 19.70 >.
$$
As in Section \ref{sec:heatexample}, the standard deviation parameters($\tau_{\sigma}, \tau_{t_{p}}$) are given half-$t$ prior distributions (with degrees of freedom $=4$).

The models were fitted using the {\tt rstan} \citep{rstan} package for {\tt R} \citep{r}, which implements a variant of Hamiltonian Monte Carlo (HMC) \citep{betancourt}.  Four chains were run, each with 2,500 iterations after 2,500 warmup iterations.  The Gelman-Rubin potential scale reduction factor was used to provide a check for adequate mixing of the four chains \citep{gelman1992inference}.

\subsection{Simulation Experiment Results}
\label{sec:sim-results}
We use the methods outlined in Section \ref{sec:predfail} to predict the number of future within-sample failures for single and multiple groups. To evaluate these methods, we calculated each procedure's coverage probability separately for lower and upper prediction bounds.  For the single group prediction, we calculated the unconditional coverage probability for each group and then averaged the probabilities over all $G$ groups.  For the multiple groups case, we calculated the unconditional coverage probability. A $100(1-\alpha)\%$ two-sided prediction interval can be obtained by combining $100(1-\alpha_L)\%$ lower and $100(1-\alpha_U)\%$ upper prediction bounds where $\alpha_L + \alpha_U = \alpha$.  Generally, it is desirable to have $\alpha_L = \alpha_U$ so that the endpoints of the interval can be interpreted as one-sided prediction bounds. This is because in most practical prediction applications, the loss from predicting too high is different than the loss from predicting too low.  To summarize our simulation results, we compare our one-sided coverage probabilities with nominal values (90\% and 95\%) to assess the adequacy of the prediction procedure.

\begin{figure}[t!]
  \includegraphics[width=1\textwidth]{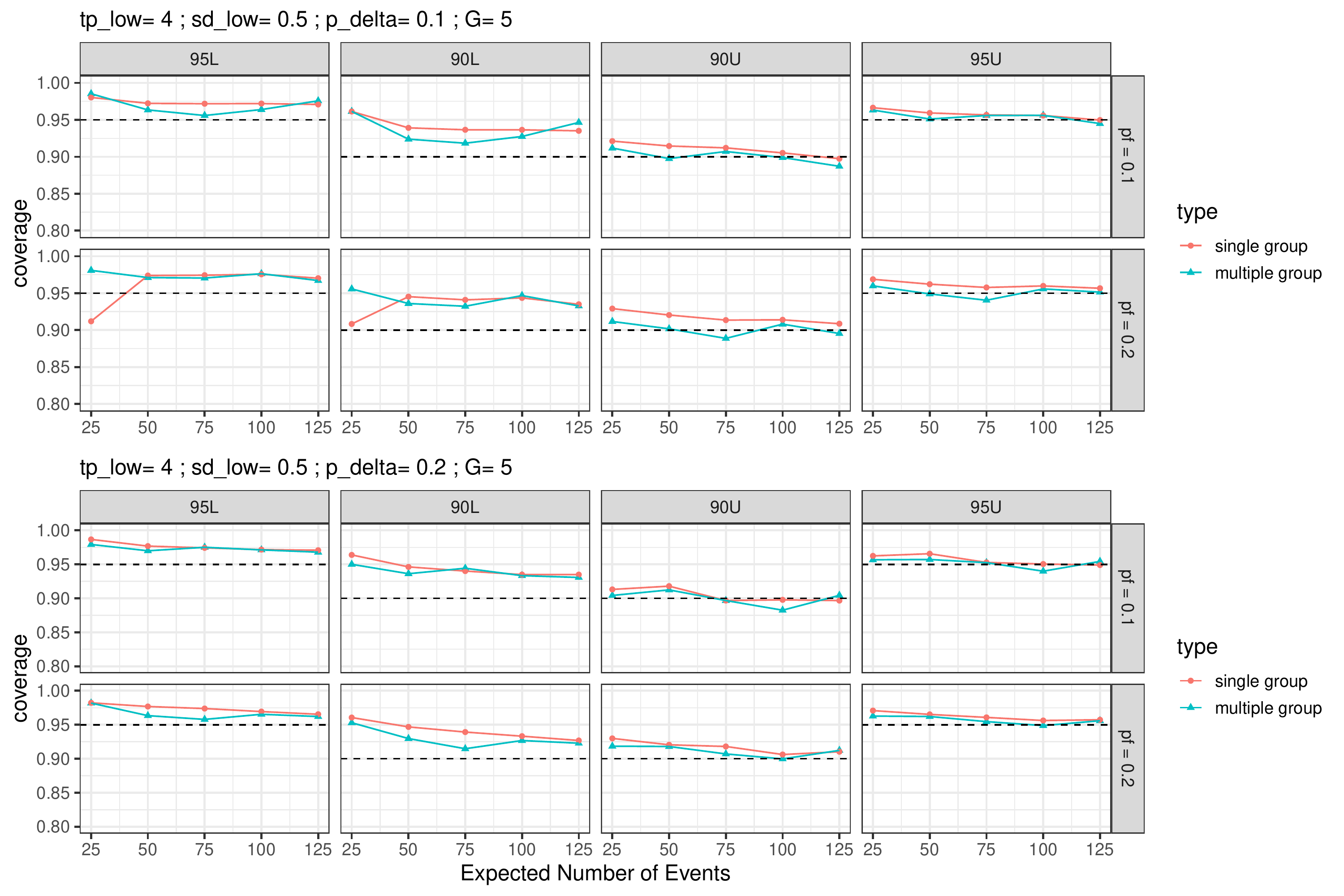}
  \caption{\footnotesize Coverage probabilities for $G=5$, $t_{p_g} \sim \text{Lognormal} < 4,8 >$, and $\sigma_g \sim \text{Lognormal} <0.50, 0.75 > $.  Top: $p_\delta = 0.10$.  Bottom: $p_\delta = 0.20$. The dashed lines correspond to the nominal coverage probability.}
  \label{f:sim1}
\end{figure}

\begin{figure}[t!]
  \includegraphics[width=1\textwidth]{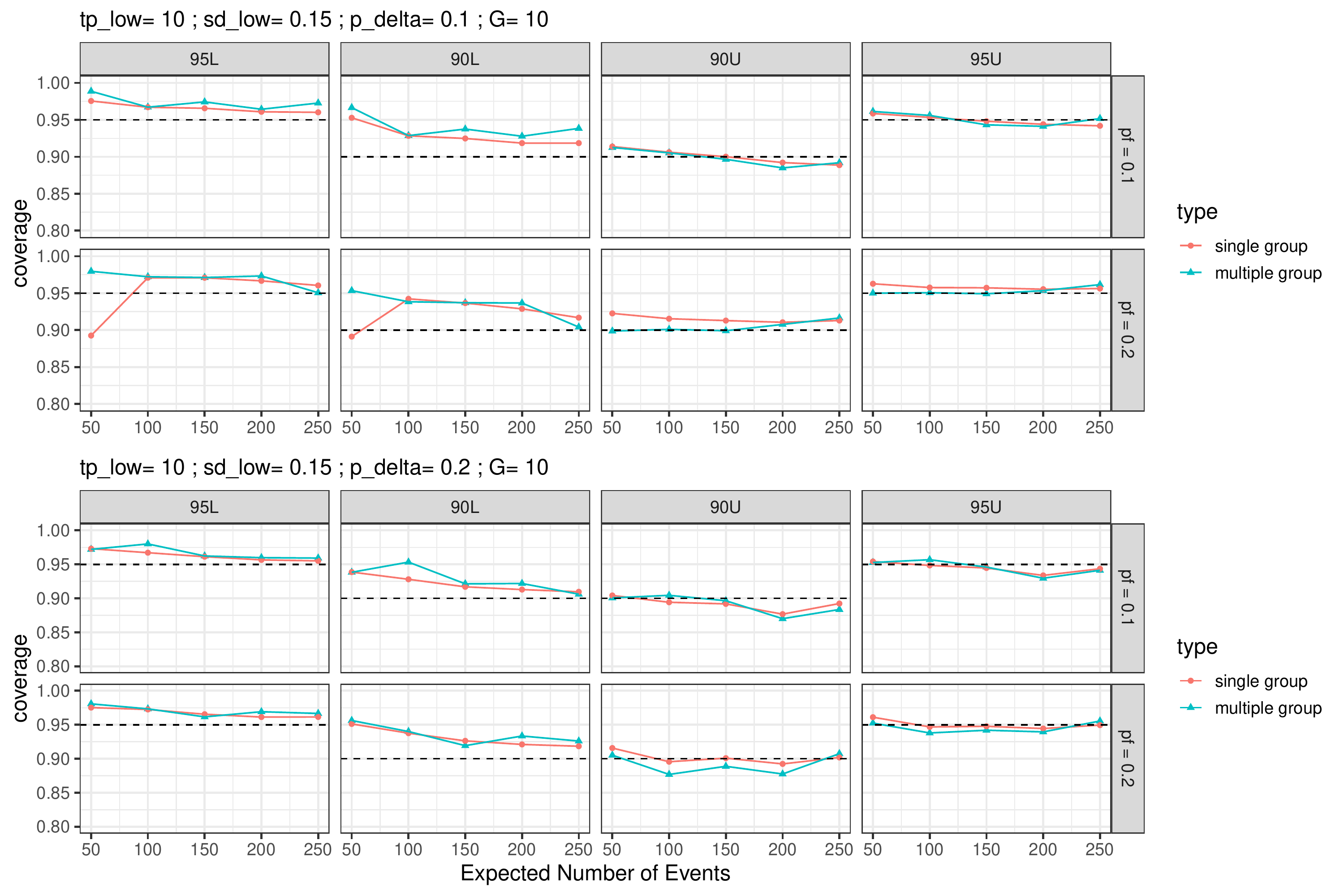}
  \caption{\footnotesize Coverage probabilities for $G=10$, $t_{p_g} \sim \text{Lognormal}< 10 ,14 >$, and $\sigma_g \sim \text{Lognormal}< 0.15, 0.40 >$.  Top: $p_\delta = 0.10$.  Bottom: $p_\delta = 0.20$. The dashed lines indicate the nominal coverage probability.}
  \label{f:sim2}
\end{figure}

Figures \ref{f:sim1} and \ref{f:sim2} present a few of the most interesting and informative graphical displays of our simulation results. Both figures show lower ($L$) and upper ($U$) coverage probabilities at the one-sided 90\% and 95\% level.  The dashed line indicates the nominal coverage level.  Both single group (circle) and multiple group (triangle) predictions are plotted. Plots for the other factor-level combinations are given in the supplementary material. Some observations from the simulation results are:
\begin{itemize}
\item The coverage probabilities for the upper one-sided prediction intervals are closer to the nominal coverage level when compared with that for the lower one-sided prediction bounds.  This is because the probability of a unit failing is small (e.g., 0.10 or 0.20), which means the lower coverage interval has a higher chance of missing an observed failure as opposed to an upper coverage bound.  
\item The multiple group predictions have over-coverage as opposed to under-coverage, especially for the lower one-side prediction bounds. For single group predictions, especially when there is a small amount of information (see top left of Figures \ref{f:sim1} and \ref{f:sim2}), predictions can perform poorly for a given subpopulation as opposed to a prediction for the entire population where there is more information. However, once the number of expected failures increases (to around 10 failures per group), performance improves. 

\item The factor levels for the Weibull parameters do not have a large effect on the coverage probabilities. Increasing $p_\delta$ from 0.10 to 0.20 does, however, move coverage closer to the nominal level.  This is due to predictions being made over a wider time window.

\item The difference between the coverage probability and the nominal one is larger when the expected number of events is small.  This is because discreteness is a greater issue when the number of events in the future is also small.  

\item Due to the hierarchical model, we can make predictions for all groups, even the ones that have no failures, by borrowing information from the groups that do have failures.
\end{itemize}

\subsection{Unbalanced Simulation Results Design}
Lastly, we extended the simulation study to a more general case where the expected number of failure values are unequal across the $G$ groups.  For this small study, we set $G=5$. We fixed the expected number of events for one group and let the expected number of events vary in the others.  Specifically, we set E$\big( r \big) = 6$ for one group, and for the remaining four we specified the levels of the expected failures to be $E\big( r \big)=4,5,6,7,8,9,10,15,20,25,30,35,40,45,50,55$.
The expected proportions of failure at the censoring time $t_c$ and the
right end of the time interval $t_w$ were, respectively, $p_{f}=0.05$ and
$p_\delta=0.5$. Using the same factors as in Section~\ref{sec:factors},
we set $t_{p_g} \sim \text{Lognormal} <4, \ 8 >$, and $\sigma_g \sim \text{Lognormal} < 0.50, \ 0.75>$.

\begin{figure}[t!]
	\centering
	\includegraphics[width=0.9\textwidth]{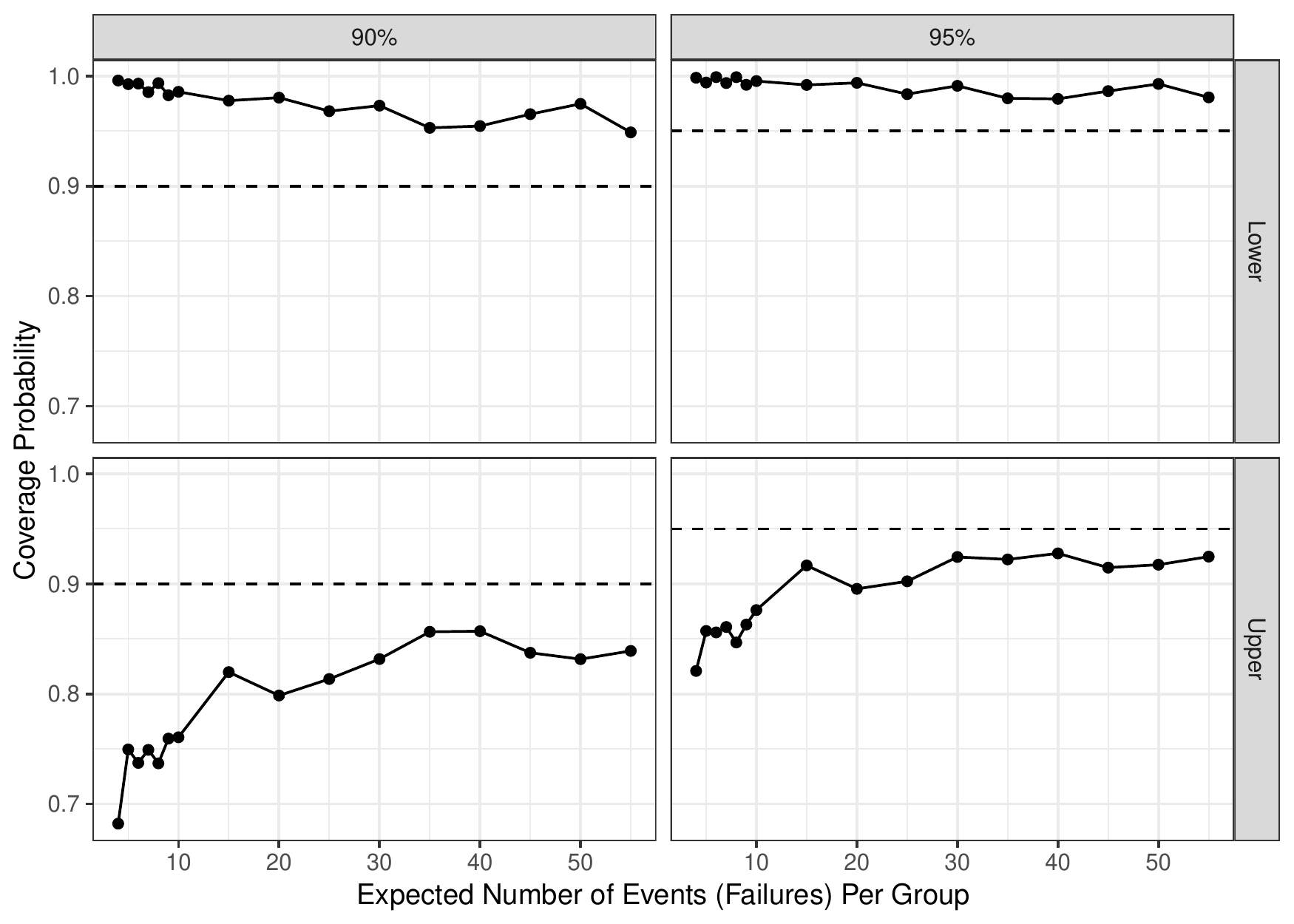}
	\caption{Coverage probability of the fixed cohort versus the expected
	number of events (failures) in the other four groups.}
	\label{fig:unbalanced-simulation}
\end{figure}

Figure~\ref{fig:unbalanced-simulation} shows a strong increasing trend toward
the nominal coverage in the upper bounds and a moderate decreasing trend toward
the nominal coverage in the lower bounds for the fixed group.
Although the expected number of events does not change in the fixed group, the coverage probability improves as the expected number of events increases in other groups; thus, the simulation results demonstrate that the hierarchical model helps to improve the prediction intervals by borrowing strength from other groups.
This feature is especially helpful when a particular group has a small sample size.

\section{Conclusions and Areas for Further Research}
\label{sec:conclusion}
This article introduces a new approach for making within-sample predictions for a population of products or systems with heterogeneous subpopulations.  The methodology is applicable to single log-location-scale families of distributions, as well as competing risk models, such as the GLFP model.  In addition, the models can include both fixed and time-varying covariates in order to improve prediction intervals. Our hierarchical modeling approach borrows strength across subpopulations with many observed failures to improve predictions for subpopulations with few failures. Estimation was performed using Bayesian methods, which allows analysts to incorporate prior information into the model. As exemplified in the heat exchanger example, informative priors can improve prediction intervals and supplement limited data. 

 We empirically verified our prediction methods using test data from Backblaze and the Product A systems, where hold-out data were available. We also conducted a small simulation study to examine the coverage probabilities of single and multiple group prediction intervals in both balanced and unbalanced designs.  The results show that, even with a small number of observed failures, Bayesian prediction intervals can achieve frequentist nominal coverage levels with weakly informative priors.  The intervals perform the best (i.e., closer to nominal level) when constructing upper prediction bounds for multiple groups, and gain precision as the number of observed failures increases.  Because lifetime data for high reliability products is often limited by sample size, censoring, and truncation, the performance of Bayesian prediction methods in small-amount-of information settings is useful for applied researchers. Some possible extensions of our work include:
\begin{itemize}
\item We assumed exchangeability of units within a subpopulation. This assumption is due in part to ignorance about potentially important covariates.  However, there may be batch effects or varying
conditions in the Backblaze facility that impact observed failure rates.  For example, the predictions for Drive-Model 21 (see supplementary material) significantly underestimate the number of future failures between weeks 1 and 10.  Then, following week 10, Backblaze removed the majority of the remaining drives from service.  Incorporating drive status information into the model (e.g., Self-Monitoring, Analysis and Reporting Technology or S.M.A.R.T. data that are available from individual drives) or other time-varying covariates could improve prediction intervals \citep{allen2004monitoring}. 

\item Our paper focused on the prediction of future events.  In some applications, however, there could be interest in predicting the cost of future events (e.g., warranty returns). Because the Bayesian approach provides full posterior distributions once a model is fit, estimating a functional, such as a depreciation factor that varies across subpopulations, would be straightforward.
\item  As seen in Figure \ref{f:pred_fails}, Backblaze retires batches of hard drives before they fail. To avoid prediction bias, we updated the risk set weekly when predicting the number of future failures.  However, an alternative approach, needed when retirements are not directly observed, is to model product retirement. \cite{xu2015assessing} used two Weibull distributions to jointly describe failure-time and retirement-time distributions.  Incorporating a retirement distribution into our methodology would allow us to predict the number of future within-sample failures over longer time intervals and to adapt our methodology to applications that have unobserved retirements.
\item In some cases, warranty claims are not immediately reported after a unit fails.  This can lead to spikes of warranty claims as the warranty contract expiration date approaches.  As seen in Figure \ref{pred-prodA}, both the proportional hazard and cumulative damage model underestimate the number of returns in June and July.  This could be due to delayed reporting.  Including such a factor in the model could improve prediction performance. 
\end{itemize}

\begingroup
	\setlength{\bibsep}{12pt}
	\linespread{1}\selectfont
	\bibliographystyle{apalike}
	\bibliography{./sample}  
\endgroup

\end{document}